\documentclass[preprint,showpacs,prl]{revtex4}

\usepackage{epsfig}
\usepackage{amsmath}
\usepackage{graphicx}
\usepackage{dcolumn}
\usepackage{amsmath}
\usepackage{latexsym}

\begin{document}

\title{The signature of a double quantum-dot structure in the I-V characteristics of a complex system}
\author{L. Bitton}
\author{D. Radovsky}
\author{A. Cohen}
\author{A. Frydman}
\author{R. Berkovits}
\address{The Minerva Center, The Department of Physics, Bar Ilan University, Ramat Gan 52900, Israel}

\begin{abstract}
We demonstrate that by carefully analyzing the temperature dependent
characteristics of the I-V measurements for a given complex system
it is possible to determine whether it is composed of a single,
double or multiple quantum-dot structure. Our approach is based on
the orthodox theory for a double-dot case and is capable of
simulating I-V characteristics of systems with any resistance and
capacitance values and for temperatures corresponding to thermal
energies larger than the dot level spacing. We compare I-V
characteristics of single-dot and double-dot systems and show that
for a given measured I-V curve considering the possibility of a
second dot is equivalent to decreasing the fit temperature. Thus,
our method allows one to gain information about the structure of an
experimental system based on an I-V measurement.

\end{abstract}

\pacs{73.63.Kv; 73.23.Hk}

\date{\today}

\maketitle

Much of the study of charging effects in quantum systems has focused
on a single-dot system, in which a metallic island is coupled to two
metallic leads via tunnel junctions \cite{QD}. If the electron
density is large enough so that the discreet energy level spacing is
negligible compared to the other energy scales, the relevant scales
are the thermal energy, $E_{T}=K_{B}T$,and the charging energy,
$E_{C}=\frac{e^{2}}{C}$, where C is the capacitance of the dot. For
low temperatures and small bias ($E_{T}<eV<E_{C}$), no current can
flow through the dot leading to a Coulomb blockade in the I-V
characteristic. An effective method to treat such a system is the
orthodox theory \cite{orthodox, orthodox1}. In this framework the
quantum dot is represented by a Double Barrier Tunneling Junction
(DBTJ) in which each junction is modeled as a resistor and a
capacitor connected in parallel. For a given voltage, $V$, a
distribution function, $\rho$, determines the probability at time
$t$ for the system to have $N$ extra electrons on the island.
Solving the master equation for $\rho$ enables one to derive the
current-voltage relation as a function of the tunneling rates. The
orthodox model assumes that the tunneling events are sequential and
an electron loses its phase coherence during a tunnel process, thus,
quantum corrections are not taken into account \cite{orthodox,
orthodox1}.

This approach has been very successful in analyzing the behavior of
a single quantum dot. In many experimental systems, however, the
exact structure is unknown. Even if a sample exhibits
Coulomb-blockade-like features, one can not always be certain that
only one simple quantum dot is involved in the transport process. An
example for such a system is demonstrated in figure 1 which shows
the I-V characteristic of a Co nanowire, $10 \mu m$ long and $200
nm$ wide \cite{daniel}. The wire was evaporated on a step-edge
structure \cite{prober} and was allowed to oxidize in atmosphere. It
is seen that the I-V curve shows Coulomb-blockade-like behavior,
presumably, due to the oxidation that gives rise to the formation of
metallic islands separated by nanoconstrictions. However, a-priori
it is impossible to know whether the structure consists of one
dominant quantum dot, two dots or perhaps even more. A similar
situation occurs in many experimental configurations and often it is
desirable to find a way to determine the exact structure of the
sample. In this paper we show that it is, in principle, possible to
fit given I-V curves to both single and double-dot configurations
using the orthodox model when the capacitance and resistance of the
barriers as well as the temperature are treated as fit parameters.
Nevertheless, since the temperature is usually well controlled in
the experimental set up, it is possible to determine whether a
system contains one or more quantum dots based on I-V measurements
at a given temperature. Most research studying the transport through
double-dot system have taken into account the discrete energy
spectrum in the dots \cite{double-dot1,double-dot2,double-dot3}. In
our approach, the more common case for metallic dots, i.e. discrete
energy level spacing much smaller than $K_{B}T$, is considered.

\begin{figure}
{\epsfxsize=5 in \epsffile{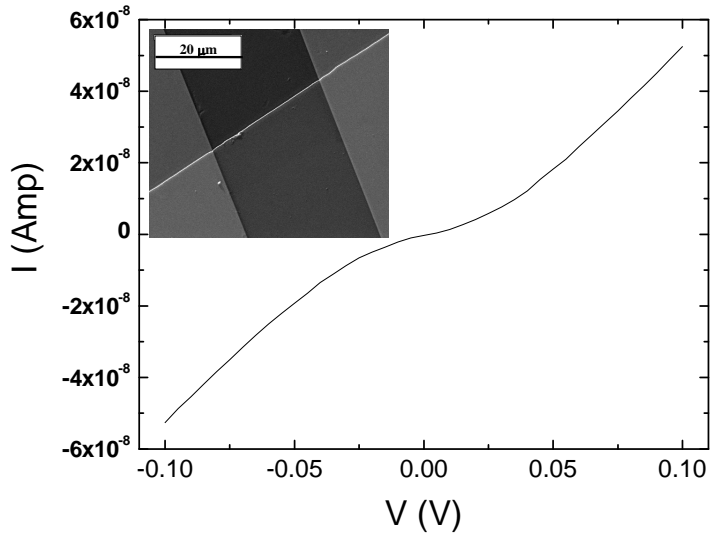}} \vspace{-2cm} \caption{\small
I-V characteristic at T=10K of a Co wire evaporated on a step-edge.
The inset shows an SEM picture of the wire (white line) grown
between two large Co electrodes (gray pads).} \label{fig_wire}
\end{figure}

The orthodox theory solution to the single dot (DBTJ) case is well
known \cite{orthodox, orthodox1}. Here we shall describe the
double-dot case which contains three tunnel junctions and is named
Triple Barrier Tunnel Junction (TBTJ). As in the DBTJ case, we
assume that the tunneling coefficient is low and tunneling events
are sequential and non-coherent, hence quantum interference effects
are neglected. In our TBTJ model each of the three junctions i=1,2,3
is characterized by a tunneling resistance $R_{i}$ and a capacitance
$C_{i}$ (see Fig. 2). The ``state'' of the system is determined by
the number of excess electrons on each grain $(n_{1},n_{2})$.
Similar to the case of DBTJ for T=0 the distribution function is
sharply peaked for each voltage around a most probable state
$(n^*_{1},n^*_{2})$. However, in contrast to the DBTJ case, one
cannot determine this state analytically and a more complex method
is required. Previous calculations on I-V characteristics of TBTJ
systems \cite{oded,
 oded1} overcame this difficulty by restricting the junction
parameters (the resistance between the dots was assumed to be much
higher dot-lead resistances), and the temperature was taken to be
zero. We suggest an approach that allows one to calculate the
distribution function in a general case without making any
assumptions on the system. Moreover we calculate the probability
value for any state, thus we can simulate the I-V curves for any
given temperature.

\begin{figure}
{\vspace{-4cm} \epsfxsize=5 in \epsffile{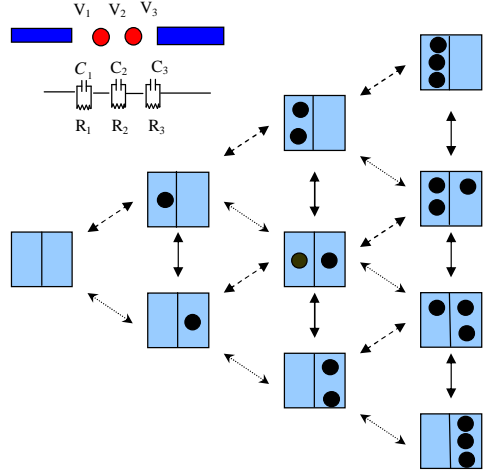}} \vspace{-4cm}
\caption{\small A schematic illustration of a system through which
three electrons are allowed to pass. Each box is divided to two
parts where each denotes one dot. Each dot is allowed to contain 0-3
electrons. The arrows indicate a permitted tunneling between states.
For the current calculation one has to sum over the transitions for
a specific kind of arrow which are related to a specific junction:
middle barrier (solid line), left barrier (dashed line) and right
barrier (dotted line). The insert shows a schematic representation
of a system composed of three tunnel junctions coupled in series.
Each junction is characterized by a capacitance and resistance. }
\label{fig_cur_trajectories}
\end{figure}
Applying voltage to a TBTJ system causes three different voltage
drops $V_{i}$ across the tunnel junctions i:

\begin{eqnarray}
 &&V_{1}=\frac{eC_{2}C_{3}V}{(C_{1}C_{2}+C_{2}C_{3}+C_{3}C_{1})},\nonumber\\
&& V_{2}=\frac{eC_{1}C_{3}V}{(C_{1}C_{2}+C_{2}C_{3}+C_{3}C_{1})},\\
&&V_{3}=\frac{eC_{1}C_{2}V}{(C_{1}C_{2}+C_{2}C_{3}+C_{3}C_{1})}.\nonumber
\end{eqnarray}

Accordingly, six tunneling rates have to be considered. Each
tunneling rate depends on the energy difference before and after the
tunneling event and on the resistance of the relevant junction. The
tunneling can be derived from Fermi's Golden rule:

\begin{equation}
 \Gamma^{i}_{\pm k}=\frac{\Delta(E^{i}_{\pm k})}{e^{2}R_{i}(1-exp(-\Delta(E^{i}_{\pm k})/k_{\beta}T))},
\end{equation}
where  $\Gamma_{\pm k}^{i} $ is the electron tunneling rate on (+)
or off (-) dot k across junction i. The energy differences,
$E^{i}_{\pm k}$  can be derived by subtracting the electrostatic
energy on the island after the tunneling event from that before the
tunneling adding the gain in energy due to the voltage drop. The
total energy differences due to tunneling of electrons from the dots
to the leads are given by:

\begin{eqnarray}
\Delta E^{1}_{\pm 1}=\frac{(C_{2}+C_{3})[(Q_{1}\pm
e)^{2}-Q_{1}^{2}]}{2(C_{1}C_{2}+C_{2}C_{3}+C_{3}C_{1})}
\mp V_{1}, \\
\Delta E^{3}_{\pm 2}=\frac{(C_{1}+C_{2})[(Q_{2}\pm
e)^{2}-Q_{2}^{2}]}{2(C_{1}C_{2}+C_{2}C_{3}+C_{3}C_{1})}\mp V_{3},
\end{eqnarray}
where $Q_{1}=en_{1}, Q_{2}=en_{2} $.

Taking into account the electrostatic energy differences of both the
dots the energy difference due to tunneling of electrons from one
dot to another is given by:

\begin{eqnarray}
 &&\Delta E^{2}_{\pm 1}=\frac{(C_{2}+C_{3})[(Q_{1}\pm
e)^{2}-Q_{1}^{2}]}{2(C_{1}C_{2}+C_{2}C_{3}+C_{3}C_{1})}\nonumber\\
&&+\frac{(C_{1}+C_{2})[(Q_{2}\pm
e)^{2}-Q_{2}^{2}]}{2(C_{1}C_{2}+C_{2}C_{3}+C_{3}C_{1})}\mp
V_{3}.\nonumber\\
\end{eqnarray}

For the TBTJ $\rho$ is the probability to have $n_{1}$ and $n_{2}$
electrons on the first and second grain respectively, hence the
master equation in a double-dot system is:

\begin{eqnarray}
{\frac{\partial \rho(V,n_{1},n_{2},t)}{\partial t}}&=& \rho(V,n_{1}-1,n_{2},t)\Gamma^{1}_{+1}+\rho(V,n_{1}+1,n_{2},t)\Gamma^{1}_{-1}\\
&&+\rho(V,n_{1},n_{2}-1,t)\Gamma^{3}_{+2}+\rho(V,n_{1},n_{2}+1,t)\Gamma^{3}_{-2}\nonumber\\
&&+\rho(V,n_{1}-1,n_{2}+1,t)\Gamma^{2}_{+1}+\rho(V,n_{1}+1,n_{2}-1,t)\Gamma^{2}_{-1}\nonumber\\
&&-\rho(V,n_{1},n_{2},t)[\Gamma^{1}_{+1}+\Gamma^{1}_{-1}+\Gamma^{2}_{+1}+\Gamma^{2}_{-1}+\Gamma^{3}_{+2}+\Gamma^{3}_{-2}].\nonumber
\end{eqnarray}

In order to find the steady state solution of the distribution
function we take ${\frac{\partial \rho}{\partial t}}=0 $. Solving
this equation requires a constraint on the number of electrons
permitted to pass through the system. For a specific number of
electrons $n_{e}$, there exists a specific number of states, $N$. In
Fig. 2 we show a schematic drawing of the transitions between the
charge configuration of a double-dot system allowing the addition of
up to three electrons to the system.

For each state we manipulate the master equation, thus achieving a
system of $N$ linear equations, where $N$ is the number of states.
This system is described by the formula:

\begin{equation}
\Gamma\cdot\vec{\rho}=0,
\end{equation}
where $\Gamma $  is a rate matrix and $\vec{\rho} $  is the states
vector. The sum over all the probabilities should be one. For
simplicity we add the normalization condition in the last row of the
rate matrix. Thus, the previous equation takes the form:

\begin{equation}
\left[
\begin {array}{ccc}
-(\Gamma^{1}_{+1}+\Gamma^{3}_{+2})&\Gamma^{3}_{-2}&\Gamma^{1}_{-1}\ldots\\
\noalign{\medskip}\Gamma^{3}_{+2}&-(\Gamma^{3}_{+2}+\Gamma^{1}_{+1}+\Gamma^{2}_{+1}+\Gamma^{3}_{-2})&\Gamma^{2}_{-1}\ldots \\
\noalign{\medskip} \Gamma^{1}_{+1}&\Gamma^{2}_{+1}&-(\Gamma^{1}_{+1}+\Gamma^{2}_{-1}+\Gamma^{1}_{-1}+\Gamma^{3}_{+2}) \ldots\\
\noalign{\medskip} \vdots & \ddots\\
\noalign{\medskip} 1&1&1 \ldots \\
\end {array}
\right] \left[
\begin {array}{ccc}
<0,0>\\
\noalign{\medskip} <0,1>\\
\noalign{\medskip} <1,0>\\
\noalign{\medskip} \vdots \\
\noalign{\medskip} <n_{e},0>\\
\end {array}
 \right]= \left[
\begin {array}{ccc}
0\\
\noalign{\medskip} 0\\
\noalign{\medskip} 0\\
\noalign{\medskip} \vdots\\
\noalign{\medskip} 1\\
\end {array}
 \right]
\end{equation}

\begin{figure}
{\epsfxsize=4 in \epsffile{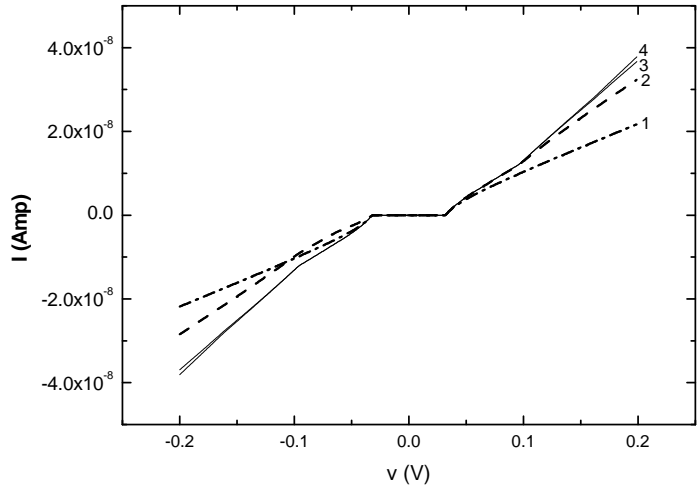}} \vspace{1cm}
\caption{\small  I-V curves for the TBTJ model T=0K,
$C_{1}=C_{2}=C_{3}=5 \cdot 10^{-18} F$,  $R_{1}=R_{2}=R_{3}=1 \cdot
10^{6} \Omega$. Each number in the graph indicates the number of
electrons passing through the system. Curves 3 and 4 which
corresponds to three and four electrons are, for all practical
purposes, identical (at least in the range of the Coulomb
Blockade).} \label{fig_cur_trajectories}
\end{figure}

By solving this numerically we obtain the distribution function and
the vector $ \vec{\rho}$. The current for a given voltage is derived
by summing over all the possibilities for an electron to pass
through a certain junction:

\begin{equation}
I(V)=\sum_{k,i,n_{1},n_{2}}C^{i}_{\pm
k}\cdot\rho(V,n_{1},n_{2})\cdot\Gamma^{i}_{\pm k}(n_{1},n_{2}),
\end{equation}
where  $ C^{i}_{\pm k}= \pm$ is determined by the direction of the
tunneling.

The results presented in this paper are obtained for a system of up
to $4$ electrons tunneling through the dot. In order to verify that
this does not lead to a considerable loss of information we plot the
numerical I-V characteristics of a typical TBTJ for $1$ to $4$
tunneling electrons in Fig. 3. It can be seen that the curves for
$3$ and $4$ electrons practically coincide for the relevant voltage
regime, hence we conclude that further increasing the number of
electrons would not have a noticeable effect on the I-V
characteristics.

Once more than two barriers are considered the system's parameter
space (i.e., possible values of $R_{i},C_{i}$) becomes large. It is
helpful then to gain some information out of the general properties
of the I-V curve. In this paper we shall concentrate on the simplest
case of symmetric I-V curves with no prominent staircase features,
which are surprisingly common in the experiments that are discussed
later. Since in the Coulomb blockade range the case for which
$R_{1},C_{1}$ differs much from $R_{3},C_{3}$ results in a
non-symmetric I-V curve and in some cases leads to staircases with
different widths and heights, we will only consider cases for which
$R_{1},C_{1}$ is similar to $R_{3},C_{3}$. Another crucial factor
for the I-V characteristics are the parameters of the middle barrier
($R_{2},C_{2}$). Choosing the parameters of the middle barrier
significantly different than the other barriers results in I-V
curves that show more prominent staircase structure. In Fig. 4. we
compare the I-V curves for the three cases discussed above. The
first case in which the parameters of the barriers are significantly
different shows a pronounced non-symmetric I-V curve. In the second
case, for which $R_{1},C_{1}=R_{3},C_{3} \ne R_{2},C_{2}$, the I-V
curve shows prominent staircase jumps. On the other hand for the
case in which all barriers are equal the I-V curve is symmetric and
smooth.

\begin{figure}
{\epsfxsize=4 in \epsffile{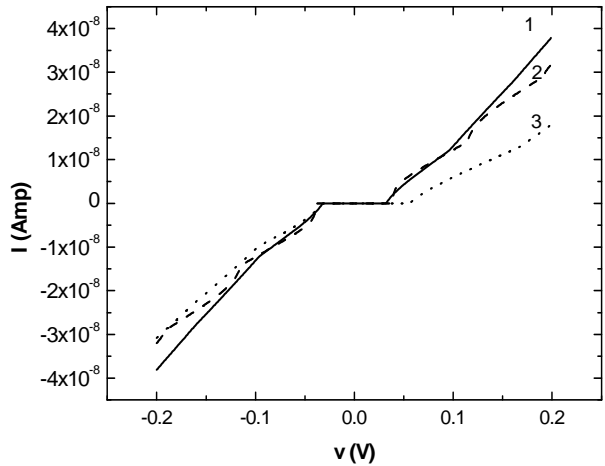}} \vspace{-0cm}
\caption{\small Theoretical I-V curves at T=0k for three cases.
Curve $1$ is the curve obtained for a pure symmetric system with the
parameters $C_{1}=C_{2}=C_{3}=5 \cdot 10^{-18} F$,
$R_{1}=R_{2}=R_{3}=1 \cdot 10^{6} \Omega$. Curve $2$ was obtained on
a system in which the parameters of middle barrier differ from rest
barriers: $C_{1}=C_{3}=7 \cdot 10^{-18} F$, $R_{1}=R_{3}=1 \cdot
10^{5}, C_{2}=3\cdot 10^{-18}, R_{2}=4 \cdot 10^{6}\Omega$. This
results in prominent staircases. Curve $3$ shows non-symmetric
behavior where the dot-lead barriers are not equal: $C_{1}=C_{2}=5
\cdot 10^{-18} F$, $R_{1}=R_{2}=1 \cdot 10^{6}, C_{3}=2\cdot
10^{-18}\Omega, R_{3}=2 \cdot 10^{6}\Omega$}
\label{fig_cur_trajectories}
\end{figure}

We demonstrate the effectiveness of our analysis in determining the
structure of a complex system by applying it to Co nanowires such as
that depicted in Fig. 1. Fig. 5 shows the experimental results and
the numerical fits of I-V characteristics of a typical nanowire
taken at different temperatures. We used the orthodox theory to fit
these data using DBTJ and TBTJ models. For the DBTJ we were able to
obtain reasonable fits only using much higher temperatures than
those of the experiment. Moreover, the ratio between the measured
and calculated temperatures increased with $T$. Using our TBTJ
calculation we were able to fit the data using the correct
measurement temperatures \cite{rem2}. This remarkable agreement for
the different temperatures reinforces our confidence in the validity
of the two-dot model to this experimental system. The diameters of
the metallic islands according to our fit are found to be $\sim 30
{\rm nm}$. This is a reasonable value since it is close to the width
of the wire.

\begin{figure}
{\epsfxsize=6.5 in \epsffile{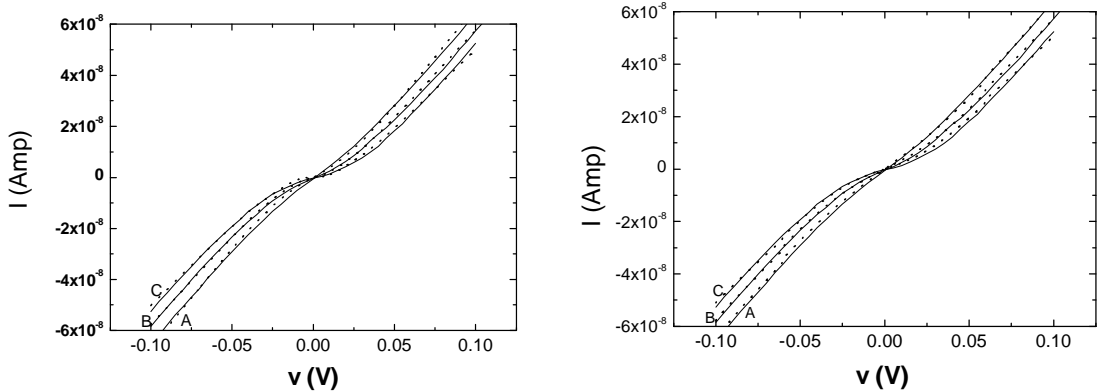}} \vspace{-4cm}
\caption{\small DBTJ (right) and TBTJ (left) theoretical fits
(dashed lines) to experimental I-V experimental curves (solid lines)
for Co nanowire samples measured at 10K (A), 20K (B), and 30K (C).
The fitting parameters for DBTJ: $R_1=R_2=7 \cdot 10^{5} \Omega$,
$C_{1}=C_{2}=4 \cdot 10^{-18} F$, T=50K (A), 65K (B), 90K (C), for
TBTJ: $R_1=R_2=R_3=4 \cdot 10^{5} \Omega$, $C_{1}=C_{2}=C_{3}=1.55
\cdot 10^{-17} F$, T=10K (A), 20K (B), 30K (C).}
\label{fig_cur_trajectories}
\end{figure}

We have applied this analysis to other wires. In some cases, even
the TBTJ model yields good fits only for temperatures much higher
than the experimental $T$. We suspect that these samples contain a
larger number of islands for which a model that takes into account
four or more tunnel junctions is required.

\begin{figure}
{\epsfxsize=6 in \epsffile{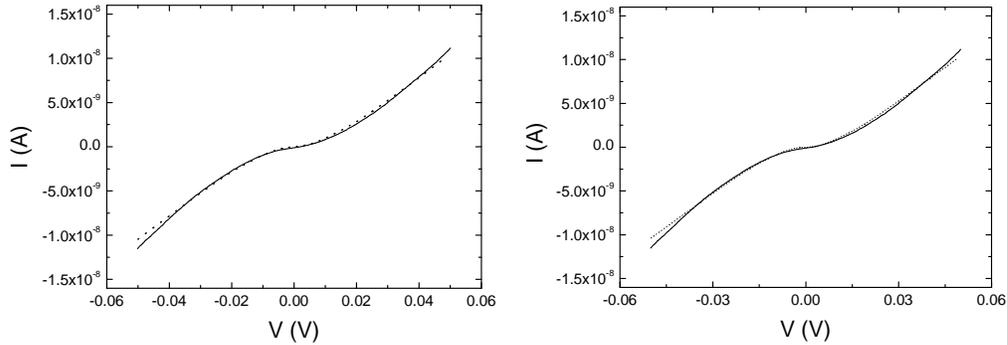}} \vspace{-4.5cm}
\caption{\small Experimental results (solid lines) and theoretical
fits (dashed lines) for the I-V characteristics of the granular
systems. The left panel is the fit for the DBTJ at
$T=30K$,$C_{1}=C_{2}=5 \cdot 10^{-18} F$, $R_{1}=R_{2}=1.55 \cdot
10^{6} \Omega$. The right panel is for the TBTJ at $T=6K$,
$C_{1}=C_{2}=C_{3}=2.8 \cdot 10^{-17} F$, $R_{1}=R_{2}=R_{3}=1 \cdot
10^{6} \Omega$  } \label{fig_cur_trajectories}
\end{figure}

Another complex system in which this analysis method may prove
useful is a disordered granular sample. In such a system the
geometrical structure may be known but the precise elements that
dominate the transport are unidentified. We  applied our analysis to
a $400 \time 400 {\rm nm}$ sample of $20-40$nm grains placed on an
insulating substrate and separated by a few nm of an insulating
matrix. Though the sample contains about $400$ grains, not all of
them participate in the transport due to the hopping nature of the
electric conductivity. Actually, it has been shown \cite{haggai}
that for low temperatures the transport is dominated by hopping
through one or two grains. I-V characteristics of such a system show
highly non-ohmic behavior that can be interpreted as signs for
charging effects. Fig. 6. shows an I-V curve taken at $T=6K$
together with fits to DBTJ and TBTJ models. Again, we were not able
to fit the results to a single dot system without increasing the
fitting temperature considerably. Using the TBTJ model, on the other
hand, we were able to obtain a very good fit for $T=6K$. From the
fitting parameters we find that the diameter of the grains is about
40 nm, which is in accordance with AFM pictures obtained on these
systems.

Both examples demonstrate that for a given experimental I-V curve,
using the TBTJ model had a similar effect to that of using a DBTJ
with higher temperature. This can be expected since the voltage drop
in the Coulomb blockade regime is divided to the contributions of
the two dots, each one contributing an energy which has to be
compared to $K_{B}T$. Thus, the measurement temperature can be an
important tool for determining the number of dots in a complex
Coulomb blockade system. For further confirmation of the analysis it
is recommendable to acquire I-V curves for different temperatures
and repeat the procedure as demonstrated in Fig. 5.

\begin{figure}
{\epsfxsize=7 in \epsffile{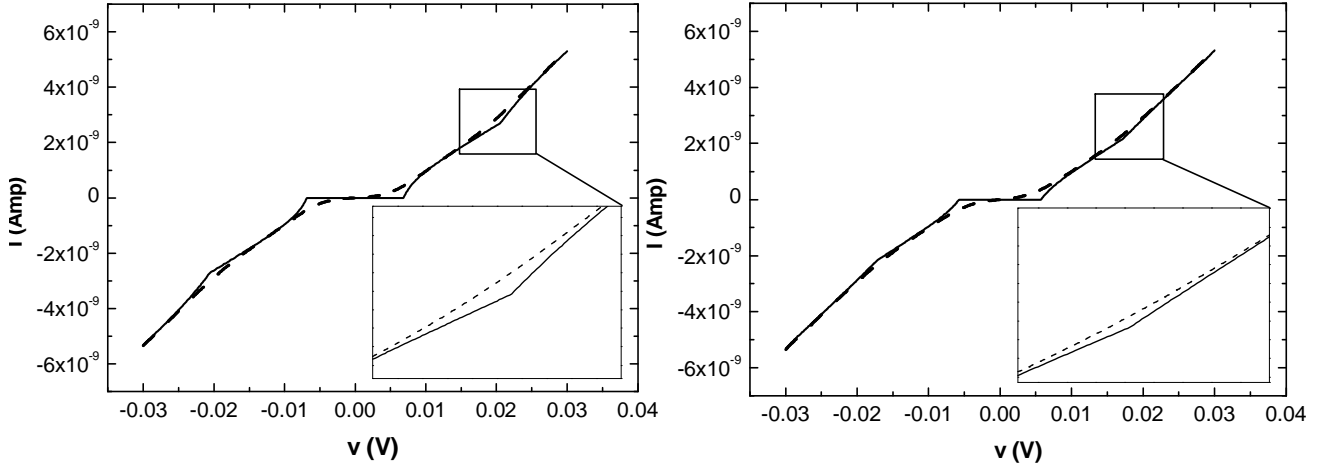}} \vspace{-2cm} \caption{\small
Theoretical results for TBTJ at T=0K (solid lines) and at T=6K
(dotted lines). The left panel is at $C_{1}=C_{3}=2.8 \cdot 10^{-17}
F$, $R_{1}=R_{3}=5 \cdot 10^{5} \Omega, R_{2}=2 \cdot 10^{6}\Omega,
C_{2}=2 \cdot 10_{-17}$. The right panel is for the total symmetric
case: $C_{1}=C_{2}=C_{3}=2.8 \cdot 10^{-17} F$, $R_{1}=R_{2}=R_{3}=1
\cdot 10^{6} \Omega$. One can notice that the staircases in the left
panel are more prominent than in right panel and the dotted lines at
both cases smears out the staircases.) }
\label{fig_cur_trajectories}
\end{figure}

As previously discussed, since the typical experimental I-V curves
considered in this paper were symmetric and had no pronounced
staircases, we choose all barrier parameters to be equal in the
fitting procedure. The assumption that all $R_{i}$s and $C_{i}$s in
the experimental structures are identical is clearly unrealistic.
Nevertheless, we find that a finite temperature smears the
difference between barriers which exhibit similar parameters. The
effect of temperature is demonstrated in Fig. 7 where the TBTJ I-V
curves for two different values of the middle barrier parameters
($R_{2},C_{2} = R_{1},C_{1} = R_{3},C_{3}$ and $R_{2},C_{2} \ne
R_{1},C_{1} = R_{3},C_{3}$) at zero temperature and at $T=6K$ are
presented. At zero temperature there is a clear staircase structure
for the latter case. At $T=6K$, on the other hand, the staircase
structure is smeared and the I-V curve is similar for both identical
and non-identical barriers. Hence, measured I-V curves cannot
determine the precise parameters of the barriers for the
experimental realizations depicted in Figs. 5 and 6 and the best we
can do is to estimate the barrier parameters to be roughly equal.
For a more exact evaluation of the barrier parameters additional
measurements at lower temperatures are required.

In summary, we propose that by analyzing the I-V characteristics one
may identify systems which are more complicated than the
conventional single-dot-double-barrier system although no former
knowledge of the system is assumed. This enables to determine the
geometrical structure of a complex quantum system. In the current
work we have implemented our approach to identifying
two-dot-triple-barrier configuration. Future work should focus on
extending this method to apply to more complex systems and,
eventually, to find a way to determine the precise number of dots in
an experimental system. This research was supported by the Israeli
Science Foundation (grants number 326/02-3, 877/04).


\begin{references}
\bibitem{rem2} In the fit we included a slight temperature
dependence of the resistances as obtained in the experiment.
\end{references}
\end{document}